\documentclass[twocolumn,floatfix,prb]{revtex4}
\usepackage{graphicx}
\usepackage{amsmath}
\newcommand{\Ef}{$E_{\rm form}$}
\begin{document}
\title{First-principles study of thin magnetic transition-metal silicide
films on Si(001)}
\author{Hua Wu, Peter Kratzer, and Matthias Scheffler}
\affiliation{Fritz-Haber-Institut der Max-Planck-Gesellschaft,
Faradayweg 4-6, D-14195 Berlin, Germany}
\begin{abstract}
In order to combine silicon technology with the functionality of
magnetic systems, a number of  ferromagnetic (FM) materials have
been suggested for the fabrication of metal/semiconductor
heterojunctions. In this work, we present a systematic study of
several candidate materials in contact with the Si surface. We
employ density-functional theory calculations to address the
thermodynamic stability and magnetism of both pseudomorphic
CsCl-like $M$Si ($M$=Mn, Fe, Co, Ni) thin films and Heusler alloy
$M_2$MnSi ($M$=Fe, Co, Ni) films on Si(001).
Our calculations show that
Si-termination of the $M$Si films is energetically preferable
during epitaxy since it minimizes the energetic cost of broken
bonds at the surface.
Moreover, we can explain the calculated trends in thermodynamic
stability of the $M$Si thin films in terms of the $M$-Si
bond-strength and the $M$ $3d$ orbital occupation.
From our calculations, we predict that ultrathin
MnSi films are FM with sizable spin magnetic moments at the Mn
atoms, while FeSi and NiSi films are nonmagnetic. However, CoSi
films display itinerant ferromagnetism.
For the $M_2$MnSi films with Heusler-type structure, the MnSi
termination is found to have the highest thermodynamic stability.
In the FM ground state, the calculated strength of the effective
coupling between the magnetic moments of Mn atoms within the same
layer approximately scales with the measured Curie temperatures of
the bulk $M_2$MnSi compounds. In particular, the
Co$_2$MnSi/Si(001) thin film has a robust FM ground state as in
the bulk, and is found to be stable against a phase separation
into CoSi/Si(001) and MnSi/Si(001) films. Hence this material is
of possible use in FM-Si heterojunctions and deserves further
experimental investigations. \indent PACS numbers: 75.70.-i,
73.20.At, 68.35.Md
\end{abstract}
\maketitle 
\section{\bf Introduction}
Metal-semiconductor heterojunctions have received much attention
in the context of magnetoelectronics or spintronics because they
could open up the possibility to inject a spin-polarized current
from a ferromagnetic (FM) metal into a semiconductor. This is a
pre-requisite for anticipated future electronic devices making use
of spin-polarized carriers.\cite{Zutic:04}
In this paper, we present theoretical investigations of thin films
for two materials classes relevant in this context, namely
transition metal (TM) mono-silicides, $M$Si ($M$= Mn, Fe, Co, Ni),
in the CsCl crystal structure, and Heusler alloys $M_2$MnSi ($M$=
Fe, Co, Ni). The two materials classes are closely related in
their crystal structure. Pictorially, one can think of $M_2$MnSi
films as being formed by the substitution of Mn for half of the Si
atoms in each Si layer of the CsCl-like $M$Si ($M$=Fe, Co, Ni)
films. Both materials classes are of potential interest for
spintronics applications. Some Heusler alloys, like Co$_2$Mn$Z$
($Z$=Si, Ge, Sn) are ferromagnets even well above room
temperature, and are predicted by band theory to be magnetic
half-metals, i.e., the Fermi energy lies in a region of partially
occupied bands for one spin channel, while lying in a gap of the
density of states in the
other.\cite{Kubler:83,Fujii:94,Galanakis-Dederichs:02} Therefore
half-metallic Heusler alloys can in principle provide 100\%
spin-polarized carriers, and could thus serve as spin-filters in
future spintronics devices. However, also the structurally simpler
mono-silicides have a potential to be applied in spintronics
devices: Recently, we have shown that thin MnSi films on Si(001)
possess sizable magnetic moments at the Mn atoms,\cite{Wu:04}
despite the fact that bulk MnSi (in the corresponding hypothetical
CsCl crystal structure) is non-magnetic.
Moreover, calculations of CoSi in CsCl crystal structure find this
(metastable) compound to be ferromagnetic.
This motivated us to study systematically both the structural and magnetic properties of late TM mono-silicides films.
In addition, mixed TM silicides have also attracted interest,
since evidence has been given that FeSi could be made
ferromagnetic by doping with Co.
\cite{Roy:02,Manyala:04}

From the viewpoint of applications, it is highly desirable to grow
well-defined FM metallic films on the most common semiconductor,
silicon, in particular on the technologically relevant Si(001)
surface. For this reason, we concentrate in the present paper on
pseudomorphic thin films of mono-silicides and Heusler alloys on
Si(001). For epitaxial growth, the mono-silicides in CsCl-like
crystal structure are particularly attractive: We find that the
CsCl structure is a metastable phase of the mono-silicides, only
moderately higher in energy than the ground state crystal
structure, and it is closely lattice-matched with Si(001).
Moreover, such CoSi and NiSi crystals have been found to be
`supersoft' materials,\cite{Moroni:98} i.e., there is a range of
elastic deformations with very little energetic cost. The Heusler
alloys show a somewhat larger lattice-mismatch with Si(001) of
about 4\%.
Apart from good lattice-match, flat and atomically sharp
interfaces are of crucial importance for efficient spin injection.
In this context, it is noteworthy that di-silicide films have been
grown with atomically sharp interfaces to Si(111) and Si(100). The
CaF$_2$ crystal structure of di-silicides is similar to the CsCl
crystal structure of mono-silicides (it results if each second
metal site in the CsCl structure is left vacant). This suggests
that film growth with atomically sharp interface should also be
possible for the mono-silicides films. In practice, first a buffer
layer of the di-silicide is grown, followed by growth of the
mono-silicide film. With this strategy, CsCl-like
FeSi and CoSi films have already been grown on Si(111) by von
K{\"a}nel et al.\cite{vonKanel:92,vonKanel:95}

While theoretical investigations of CsCl-like $M$Si thin films on
Si(001) are scarce,\cite{Profeta:05}
a group of studies addressing the initial reaction
processes of TM adatoms with the Si substrate report that Mn, Co
and Ni adatoms prefer subsurface
sites.\cite{Dalpian:04,Wu:04,Horsfield:01,Higai:00}
Heusler alloy films have been studied experimentally mostly in
view of their application in tunnelling magneto-resistance
devices.\cite{Kammerer:03,Kammerer:04,Schmalhorst:04} Concerning
epitaxial growth on semiconductor substrates, results for thin
Co$_2$MnGe \cite{Ambrose:00} and Co$_2$MnSi \cite{Wang:05}
films on GaAs(001) have been
reported. From the theoretical side,
calculations of the Co$_2$MnSi(001)
surface,\cite{Galanakis:02,Hashemifar:05} as well as of the
interface between Co$_2$MnGe and
GaAs(001)\cite{Picozzi:03,Picozzi:03a} have been performed.

In the present paper, we identify the trends in chemical bonding,
thermodynamic stability, and magnetism of the $M$Si and $M_2$MnSi
thin films. Most importantly, our calculations predict that, in
addition to ultrathin FM MnSi/Si(001) films,\cite{Wu:04} the
CoSi/Si(001) thin films are also FM; and that Co$_2$MnSi/Si(001)
films have a robust FM ground state.

\section{\bf Computational details}
The present DFT calculations were performed using the all-electron
full-potential augmented plane-wave plus local-orbital
method.\cite{BlSc01} The generalized gradient approximation
(GGA)~\cite{PeBu96} was adopted for the exchange-correlation
potential, since it has been shown~\cite{PhBa96,Moroni:99} that
GGA gives a better description for both transition metals and
their silicides than the local-spin-density approximation. The
$M$Si or $M_2$MnSi thin films on Si(001) were modelled by a slab
consisting of eight successive Si(001) layers and the $M$Si (see
Fig. 1) or $M_2$MnSi layers (see Section \ref{Heusler}) on both
sides, in order to retain the inversion symmetry.
The GGA calculated equilibrium lattice constant (5.48 \AA~) of
bulk Si is used for the Si(001) substrate.
A supercell with about 10-11~{\AA} vacuum between the
\begin{figure}[h]
\centering \includegraphics[width=8cm]{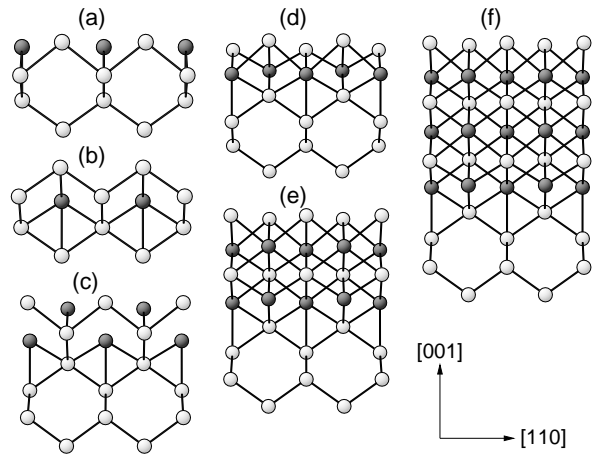} \caption{Side
view of various $M$=Mn, Fe, Co, or Ni films on Si(001) (half of
the slab), with 0.5
ML $M$ in (a) the first- or (b) second-layer interstitial sites, 1
ML $M$ (c) in a mixed layer or (d) in a Si-$M$ sandwich, or (e) 2
ML or (f) 3 ML $M$ CsCl-like sandwich structures. Black balls
represent $M$ and gray balls Si atoms. The bonds shorter than
2.65~{\AA} are shown.}
\end{figure}
slabs, and with a lateral $(1\times 1)$ periodicity~\cite{Wu:04}
(lattice constant of 3.87 \AA~)
was used. Note that $\theta$=1 ML (monolayer) coverage of $M$
refers to two $M$ adatoms per $(1\times 1)$ cell on either side of
the slab. The muffin-tin radii are chosen to be 1.11~{\AA} for Mn,
as used in our previous calculations,\cite{Wu:04} and 1.06~{\AA}
for Fe, Co, Ni, and Si, in order to avoid overlap of the
muffin-tin spheres (due to covalent bond-shortening within the TM
silicide series, as we report below) during structure relaxations.
This choice is reasonable in view of their respective atomic
sizes. The cut-off energy for the interstitial plane-wave
expansion is chosen to be 15.2 Ryd.\cite{Note} A set of
10$\times$10$\times$1 special $\bf{k}$ points is used for
integrations over the Brillouin zone of the $(1\times 1)$ surface
cell. Except for the two central Si layers in the slab,
all the $M$ and other
Si atoms are relaxed until the calculated atomic force for each of
them is smaller than 0.05 eV/{\AA}. Throughout this paper,
formation energies are given per $(1\times 1)$ cell, defined as
\begin{equation}
E_{\mathrm{form}}=(E_{\mathrm{tot}}-\sum_{i}N_i\mu_i)/2
-\gamma_{\mathrm{Si}}A,
\end{equation}
where $E_{\mathrm tot}$, $N_i$ and $\mu_i$ refer to the total
energy per $(1\times 1)$ unit cell with surface area $A$, the
number of atoms of each chemical type in the cell, and their
chemical potentials as calculated from the corresponding bulk
materials. The factor 2 in the denominator is because
the slab contains two equivalent surfaces due to the inversion symmetry.
$\gamma_{\mathrm{Si}}$=84 meV/{\AA}$^2$ is the surface
energy of the clean, $p(2 \times 2)$-reconstructed Si(001)
surface.
We note that $E_{\mathrm{form}}$ defined in this way contains 
the bulk heat of formation, as well as surface and interface contributions. The interface energy alone, which could serve as an indicator for adhesion of the films to the substrate, is not considered.
The numerical accuracy of the present calculations is
carefully checked by using higher cut-off energy and more $\bf{k}$
points. With these settings, the absolute values of
$E_{\mathrm{form}}$ are converged with respect to cut-off energy
and {\bf k}-point sampling to better than 0.1 eV. However, for the
{\em relative} stability of structures with the same composition
but different geometries and/or magnetic structures, we can give a
much stricter error estimate, only several meV, due to error
cancellation since all numbers entering the energy difference are
calculated with the same technical settings.
The degree of spin polarization at the Fermi level is quantified
from the spin-resolved density of states (DOS), which is
calculated using a finer {\bf k}-point mesh of $16 \times 16
\times 1$ in conjunction with the tetrahedron method for Brillouin
integration. We note that a more realistic assessment of spin
injection at the interface would have to consider the match in
Fermi velocities in the film and the substrate. For bulk magnets,
a spin polarization including a suitable weighting with the Fermi
velocity can be defined\cite{Mazin:99,Panguluri:03}. However, in
this work we retain the more wide-spread definition of the DOS.

\section{\bf Results and Discussion}
\subsection{\bf bulk phases of $M$Si}

Before studying the $M$Si thin films on Si(001), we briefly
discuss the bulk phases of the TM mono-silicides $M$Si ($M$=Mn,
Fe, Co, Ni). For all metal atoms discussed here, the
mono-silicides have the same bulk crystal structure, the B20
structure, whose symmetry is characterized by the $P$2$_1$3 space
group.\cite{Marel:98} Since the lattice constant of the cubic unit
cell is around 4.5~{\AA} for all these compounds, they cannot be
lattice-matched with Si(001). However, the metastable CsCl phase
calculated within DFT-GGA lies only slightly above the
ground-state $P$2$_1$3 structure in total energy, for $M$=Mn, Fe,
Co, and Ni by 0.25, 0.04, 0.42, and 0.24 eV per formula unit,
respectively. Moreover, it follows from our GGA calculations that
the equilibrium lattice constants for the metastable CsCl phases
are 2.79, 2.77, 2.78, and 2.85~{\AA}, respectively. They are
almost half the calculated lattice constant of Si (5.48~{\AA}),
and thus the lattice mismatch with Si(001) is less than 2\% for
the CsCl-like MnSi, FeSi, and CoSi, and 4\% for NiSi. These
results for $M$Si ($M$=Fe, Co) agree well with the previous
calculations by Moroni, Podloucky, and Hafner.\cite{Moroni:98}

We show in Fig. 2 the density
of states of the CsCl-like $M$Si calculated within GGA in
the nonmagnetic (NM) state. The CsCl-like FeSi and
NiSi have a low DOS at the Fermi level, which explains, within the
framework of the Stoner model of magnetism, why we find them to be
non-magnetic. In contrast, the Fermi level of the CsCl-like MnSi
lies at a falling shoulder of the $t_{2g}$ DOS. In particular, the
Fermi level of the CsCl-like CoSi lies at a steep slope of the
$e_g$ DOS, which gives rise to Stoner FM instability. This has
also been discussed by Profeta $et$ $al$.\cite{Profeta:05}
Our calculations show that the FM
ground state of CoSi has a spin moment of 0.63 $\mu_B$/Co and a
lower total energy than the NM state by 16 meV per formula unit.

Since epitaxial growth of the CsCl-like FeSi and CoSi films on
Si(111) has already been achieved by von K\"anel {\it et
al.},\cite{vonKanel:92,vonKanel:95} and given that CoSi has the
highest energy difference for the metastable phase among the
CsCl-like $M$Si ($M$=Mn,Fe,Co,Ni), we consider it likely that
growth of the CsCl-like $M$Si films on Si(001), and of the
CsCl-like MnSi and NiSi films on Si(111), can be achieved as well.
\begin{figure}
\includegraphics[width=8.5cm]{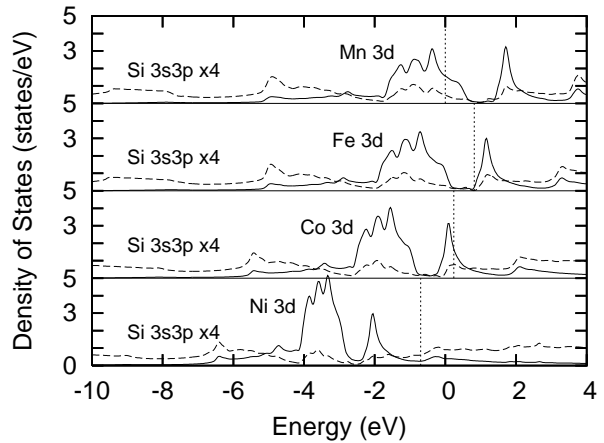}
\caption{Orbital-projected DOS of metastable CsCl-like bulk $M$Si
($M$=Mn, Fe, Co, Ni) in the non-magnetic state. The solid lines
refer to the $M$ $3d$ bands, which split into the lower-lying
$t_{2g}$ and the higher-lying $e_g$ bands.
The dashed lines refer to the Si $3s3p$ states (magnified four
times for clarity). The Fermi level of MnSi (calculated to be
11.76 eV) is used as energy zero for {\em all} plots. The Fermi
levels (vertical dotted lines) of FeSi, CoSi, and NiSi differ from
that of MnSi by 0.81, 0.23, and $-0.69$ eV, respectively.
Obviously, the shapes of those DOS are similar, and the Fermi
level shifts towards and strides over the $e_g$ states to
accommodate more and more $d$-electrons as $M$ varies from Mn
through Fe and Co to Ni. Note that, as $M$ varies from Mn to Ni,
the $M$ $3d$ bands monotonously shift down toward the Si $3s3p$
valence bands.}
\end{figure}

\subsection{\bf $M$Si thin films on Si(001)}

For various amounts of TM atoms deposited on Si(001), we
perform calculations to investigate the stable binding sites or
the (meta-)stable atomic structure of films. As seen below,
the preceding calculations
for $\theta$=0.5 ML and 1 ML are helpful to understand why the
$M$ atoms prefer subsurface sites and the Si atoms sit in the topmost layer.

We start our calculations by considering a coverage of
$\theta=$0.5 ML of metal atoms $M$, occupying either atomic sites
{\em on the surface} [cf. Fig. 1(a)] or {\em subsurface} sites
[cf. Fig. 1(b)] of Si(001). The results show that all metal
adsorbates, $M$=Mn, Fe, Co, and Ni, are generally more stable at
Si(001) subsurface than at surface sites, by about 0.1 eV per
(1$\times$1) cell for $M$=Mn, and more than 0.4 eV for $M$=Fe, Co,
or Ni, as seen in Table~\ref{TableI}. The surface adatoms $M$=Mn,
Fe, and Co have a sizable spin moment, and in Table~\ref{TableII}, the
values within the atomic muffin-tin spheres are reported.
The reduction of the spin magnetic moment of $M$ atoms
on subsurface sites is due to the increased number of $M$-Si bonds. In
particular, the magnetic moment of the subsurface Co atom is
almost completely quenched. Moreover, we find Ni atoms to have
vanishing magnetic moments both on the surface and at subsurface
sites. Note that in these $M$-Si ($M$=Mn, Fe, Co) systems, spin
\begin{table}[th]
\caption{Formation energies [in units of eV per (1$\times$1) cell]
of films in various structures depicted in Fig. 1, labelled
a)--f), relative to the clean Si(001) surface and elemental bulk
$M$=Mn, Fe, Co, or Ni. Note that the values of {\Ef} in the $M$=Mn
row are slightly different (by 0.03 eV at most) from those of our
previous calculations\protect\cite{Wu:04} given in parenthesis,
due to different values of the muffin-tin radius of Si and the
cut-off energy used.} \label{TableI}
\begin{tabular} {lcccccc} \\ \hline\hline
{\Ef}&a&b&c&d&e&f \\ \hline
Mn&0.76&0.67&0.89&0.61&--0.43&--1.55\\
&(0.77)&(0.68)&(0.90)&(0.62)&(--0.40)&(--1.53)\\
Fe&1.11&0.67&0.93&0.01&--1.71&--3.78 \\
Co&0.99&0.47&0.89&--0.44&--2.38&--4.15 \\
Ni&0.59&0.18&0.22&--0.64&--2.37&--3.46  \\ \hline \hline
\end{tabular}
\end{table}
\begin{table}[bh]
\caption{Spin magnetic moment (in unit of $\mu_B$) of $M$ atoms
within muffin-tin spheres for various structures depicted in
Fig.~1, labelled a)--d). NM Ni case is omitted. Reported for c)
are both values for the
surface and subsurface $M$ atoms, separated by a comma; and
for d) are the substitutional and interstitial $M$ atoms.}
\label{TableII}
\begin{tabular} {lcccccc} \\ \hline\hline
$m$&a&b&c&d&& \\ \hline
Mn&3.68&3.08&3.26, 2.25&2.16, 1.65&&\\
Fe&2.35&2.09&2.45, 1.94&0.11, 0.05&& \\
Co&0.95&0.03&0.45, --0.07&0.41, 0.35&& \\ \hline\hline
\end{tabular}
\end{table}
moments are also induced on the Si atoms adjacent to $M$, albeit
smaller than 0.1 $\mu_B$.

Secondly, we compare two possible atomic structures for 1~ML
coverage, the 1ML-$M$ surface mixed layer [cf. Fig. 1(c)] and the
layered Si-$M$ film [cf. Fig. 1(d)]. Our results show that the
latter is energetically more favorable than the former, by about
0.3 eV per (1$\times$1) cell for $M$=Mn and around 1.0 eV for
$M$=Fe, Co, or Ni. Next, we analyse the chemical bonding in these
systems. We start by noting that $M$ (=Mn, Fe, Co, or Ni) and Si
have almost identical electronegativity of 1.6 or 1.7, and hence
form strong covalent bonds. From Fig. 3, we see that the $M$-Si
bonds have similar covalent charge density as the Si-Si bonds.
Moreover, for all relaxed structures of the Si-$M$/Si(001)
($M$=Mn, Fe, Co) films, we find that both the substitutional $M$
(named $M$1) and the interstitial $M$ (named $M$2) each have four
$M$-Si bonds which are shorter, by 0.13~{\AA} at least, than the
sum of the $M$ and Si atomic radii, due to covalent bond
contraction. NiSi is an exception to this general trend; in
Si-Ni/Si(001) the substitutional Ni1 has four shrunk Ni-Si bonds
which are contracted by 0.08~{\AA}, and the interstitial Ni2 has
only two short Ni-Si bonds, contracted by 0.18~{\AA}. This
exceptional behavior, both the smaller Ni1-Si bond-shortening and
\begin{figure}[th]
\centering \includegraphics[width=8.5cm]{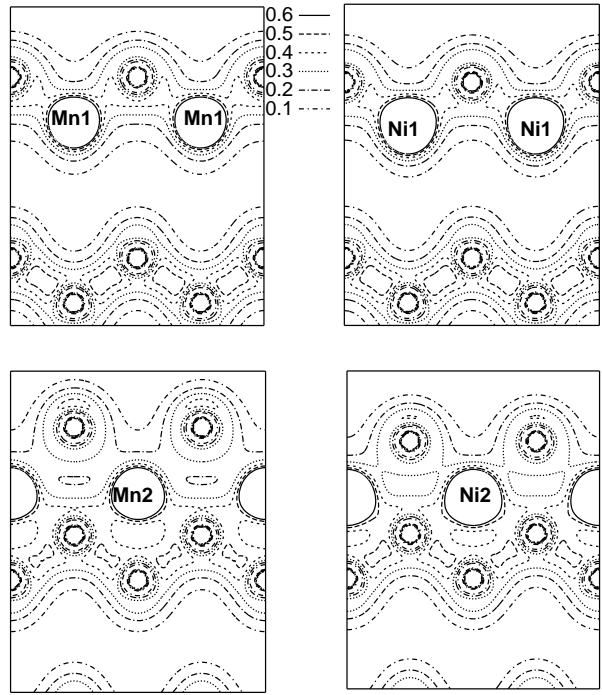}
\caption{Valence charge density in the (1$\overline1$0) plane for
1~ML Si-capped silicide films, Si-$M$/Si(001) [$M$=Mn (left
panels) or Ni (right panels), cf. Fig. 1(d)]. The cuts are chosen
to contain the substitutional-$M$1 and Si (upper row), or the
interstitial-$M$2 and Si atoms (lower row). Contour lines from 0.1
to 0.6 e/\AA~$^3$ in steps of 0.1 e/\AA~$^3$ are shown. The Mn-Si
and Ni-Si bonds have a covalent charge density as high as 0.4
e/\AA~$^3$, similar to the Si-Si bonds with 0.5 e/\AA~$^3$.}
\end{figure}
the reduced number of short Ni2-Si bonds, can be understood by
considering that the number of empty $3d$ orbitals available for
bonding with Si decreases in the TM series from Mn to Ni.
Note that the transition metal atoms are seven-fold coordinated to Si in the
natural bulk silicides $M$Si, and eight-fold coordinated in
$M$Si$_2$. Thus, the subsurface TM layer capped by a Si layer in
the Si-$M$/Si(001) films optimizes the surface covalent bonding
structure, since it allows for the optimum fourfold coordination
of the capping Si atoms, while simultaneously increasing the
coordination of the $M$ atoms (compared to {\em on surface}
adsorption). The Si-termination of the CsCl-like FeSi/Si(111) film
surface has been previously verified both experimentally and
theoretically.\cite{Walter:03} Moreover, the Si capping layer, due
to the doubled atomic density as compared with the Si(001)
substrate, displays strong buckling, 0.43, 0.57, 0.47, and
0.21~{\AA} in the Si-$M$/Si(001) film with $M$=Mn, Fe, Co, or Ni,
respectively.

Since the layered Si-$M$ film has turned out to energetically most
favorable from the above calculations, we employ the same atomic
structure to multilayered Si-$M$ [$n$(Si-Mn)] films, i.e., to the
CsCl-like $M$Si films with Si termination, as depicted in Figs.
1(e) and 1(f). As seen in columns (d), (e) and (f) of
Table~\ref{TableI}, The formation energy {\Ef}, definded according
to Eq. (1), decreases monotonously with increasing film thickness
for all CsCl-like $M$Si films. This decrease is a consequence of
the heat of formation released for each formula unit of $M$Si
formed from the elements. The onset of negative {\Ef} at
$\theta\approx 2$~ML Mn or 1 ML $M$ ($M$=Fe, Co, Ni) indicates
that the films are stable against decomposition into the clean
Si(001) surface and elemental bulk $M$.

Moreover, the thermodynamic stability of the $M$Si films increases
as $M$ varies from Mn through Fe, Co to Ni at $\theta<$2 ML. We
attribute this finding to the increasing $M$--Si bond strength:
Note that {\Ef} is calculated with reference to the clean Si(001)
surface and elemental TM bulk (see Eq. 1). Both GGA calculations
and experimental measurements agree that the cohesive energies of
Fe, Co, and Ni are very similar, and higher than that of Mn by
about 1~eV.\cite{PhBa96} Therefore the decreasing {\Ef} of the
$M$Si films at $\theta<2$~ML as $M$ varies from Mn to the later
TMs indicates that the binding energy of the $M$ atoms on Si(001)
increases more strongly so as to overcompensate the rise in the
removal energy of an $M$ atom from its bulk reservoir upon
variation of $M$ from Mn to the later TMs. Hence, the strength of
the $M$--Si bonds must increase accordingly. This trend can be
understood by observing that the $M$ $3d$ bands increasingly come
into resonance with the Si $3s3p$ valence bands due to decreasing
energy separation between them (see Fig. 2), because the $M$ $3d$
level shifts down towards the Si $3s3p$ level as the atomic number
of the transition metal increases. However, the trend is reversed
for the NiSi film at $\theta=2$~ML [see column (e) in
Table~\ref{TableI}]. For thicker $M$Si films, the order of
thermodynamic stability, quoted from low to high, changes to
$M$=Mn, Ni, Fe, Co at $\theta=3$~ML [see column (f) in
Table~\ref{TableI}]. The anomaly in the NiSi case can be explained
in terms of $M$ $3d$ orbital occupation. Since Ni has the fewest
empty $3d$ orbitals available for bonding with Si, the Ni atoms in
the NiSi film (except for the interfacial Ni) being eightfold
coordinated to Si become oversaturated. The oversaturation for
eightfold Si coordination of Ni is also reflected by the increased
lattice constant of the CsCl-like NiSi [compared with $M$Si
($M$=Mn,Fe,Co) as seen in Sec. III. A]. This interpretation is
corroborated by the experimental observation that the lattice
constant of the eightfold coordinated NiSi$_2$ is larger than that
of CoSi$_2$.

The above results are helpful to understand three experimental
observations. Firstly, pre-adsorbed Co has been found to improve
the quality of Fe films grown on Si(001).\cite{Bertoncini:99} Our
calculations show that Co--Si bonds are stronger than Fe--Si
bonds; hence the improved film quality can be explained by a CoSi
layer forming at the interface which prevents interdiffusion
between the Fe overlayer and the Si substrate. Moreover, we can
predict that Ni cannot be used for this purpose, because the
highly Si-coordinated Ni-silicide is thermodynamically less stable
than Fe-silicide, as we reported above. Hence, we conclude from
our calculations that Ni is unsuitable for a barrier layer to
suppress the intermixing between Fe and Si. Secondly, the trends
in bond strength revealed by our calculations help to explain the
structure of Heusler alloys with the chemical composition
$M_2$MnSi ($M$=Fe,Co,Ni), or more generally
$X_2YZ$,\cite{Kubler:83,Fujii:94,Galanakis-Dederichs:02} in which
$X$, $Y$ and $Z$ have a similar electronegativity and $Y$
possesses a robust magnetic moment. In these materials, so-called
full Heusler alloys, which can be considered as a (111) stacking
of layers with the sequence $Z-X-Y-X-Z-X-Y-X-Z\ldots$, it is
always the element $X$ capable of making stronger bonds to $Z$
which occurs in the layers adjacent to $Z$, while the more weakly
bonding element $Y$ has $Z$ only as its second neighbors. Together
with knowledge of the energetic positions of the atomic levels of
the $X$, $Y$, and $Z$ atoms, and thus of their relative bond
strengths, this rule can be used as heuristics in the search for
new Heusler alloys (some of which may be half-metallic FMs),
somewhat similar in spirit to the `band gap engineering' done in
semiconductor physics. Thirdly, on the basis of our results, we
can explain the observed site selectivity\cite{Burch:81} for
substitution of other TMs in the Heusler alloy
Fe$^{A}_{2}$Fe$^B$Si: The TMs to the right of Fe in the periodic
table, Co and Ni, making stronger bonds to Si than Fe itself,
substitute for Fe$^A$ to form new stronger bonds with four Si
neighbors. The earlier TMs Ti, V, Cr, Mn, however, substitute for
Fe$^B$, thus preserving the stronger Fe$^A$-Si bonds.

Next we turn to the magnetism of the $M$Si thin films on Si(001)
[$n$(Si-$M$)/Si(001)]. As a general trend in the pseudomorphic
(Si-$M$)/Si(001) films [cf. Fig. 1(d)], we find that the
substitutional $M$1 (cf. Fig. 3) has a little larger spin
moment (e.g., 2.16
$\mu_B$/Mn1) than the interstitial $M$2 (e.g., 1.65 $\mu_B$/Mn2),
as seen in Table~\ref{TableII}. This can be partly ascribed to the
number of $M$-Si bond being fewer by one for $M$1 (six-fold
coordination) than $M$2 (seven-fold coordination). First, we
describe in more detail the results for MnSi films. The
(Si-Mn)/Si(001) film is found from our calculations to be a
ferromagnetic metal with a sizable spin moment, in which the Si
atoms mediate the FM Mn-Mn coupling via hybridization between the
Si $3s3p$ and Mn $3d$ itinerant electrons. A vital role is played
by the capping Si atoms; in their absence the bare Mn film on
Si(001) is found to be antiferromagnetic (AFM).\cite{Wu:04} For
the 2(Si-Mn)/Si(001) film, our calculations also predict a FM
metallic ground state. The 3(Si-Mn)/Si(001) film is found to be
ferrimagnetic with FM (ferrimagnetic) intra (inter)-layer
coupling, as seen in Tables~\ref{TableIII} and \ref{TableIV}. The
middle Mn layer has a small spin moment of $-0.14\, \mu_B$/Mn
antiparallel to the larger one of 1.74 $\mu_B$/Mn in the
interfacial Mn layer. It mediates a superexchange ferrimagnetic
coupling between the interfacial and subsurface Mn layers. Note
that the interlayer magnetic coupling is weak in the
$n$(Si-Mn)/Si(001) thin films, e.g., the energy cost for flipping
the magnetic moments of one layer, i.e., going from FM to AFM
ordering between layers, is 8 and 10 meV/Mn in the
2(Si-Mn)/Si(001) and 3(Si-Mn)/Si(001) films, respectively.
However, the FM intralayer coupling is rather strong, as is
\begin{table}[h]
\caption{Spin magnetic moment (in unit of $\mu_B$) of atoms averaged
over one layer
[from interface layer (left) to surface layer (right)] of the
$M$Si thin films on Si(001) [cf. Figs. 1(d), 1(e) and 1(f)]
in their respective magnetic ground
states. Note that the FeSi/Si(001) films are non-magnetic, as
discussed in the text. The non-magnetic NiSi/Si(001) films are omitted.}
\label{TableIII}
\begin{tabular} {lcccccc} \\ \hline\hline
&$M$&Si&$M$&Si&$M$&Si \\ \hline
Si-Mn&1.90&--0.05 \\
Si-Fe&0.08&--0.01 \\
Si-Co&0.38&0.02 \\ \hline
2(Si-Mn)&1.90&--0.07&1.11&0.02 \\
2(Si-Fe)&0.38&--0.01&0.06&0.01 \\
2(Si-Co)&0.16&--0.01&0.55&--0 \\ \hline
3(Si-Mn)&1.74&--0.03&--0.14&0.03&--1.07&--0.04 \\
3(Si-Fe)&0.31&--0.01&0.01&--0&0.01&+0 \\
3(Si-Co)&0.38&--0.01&0.56&--0.01&0.63&--0.01 \\ \hline\hline
\end{tabular}
\end{table}
\begin{table}[h]
\caption{Total-energy difference (in units of meV per $M$ atom)
of the $n$(Si-$M$)/Si(001) ($n$=1,2,3; $M$=Mn,Fe,Co) thin
films among the ferromagnetic (FM), antiferromagnetic [AFM, either
intra- (or inter-) layered AFM marked with superscript $i$ (or $o$)],
and non-magnetic (NM) states.} \label{TableIV}
\begin{tabular} {lccccccccccccr} \\ \hline\hline
&&$n$(Si-Mn)&&&&&$n$(Si-Fe)&&&&&$n$(Si-Co)& \\ \hline
$n$&1&2&3&&&1&2&3&&&1&2&3 \\ \hline
FM&0&0&10&&&0&0&0&&&0&0&0\\
AFM&71$^i$&8$^o$&0$^o$&&&FM&0&0&&&NM&0&10$^o$\\
NM&350&188&80&&&0&5&0&&&15&17&28\\ \hline\hline
\end{tabular}
\end{table}
evident from the energy cost for flipping one of the two magnetic
moments per layer in the unit cell, i.e., going from FM to AFM
ordering within the layers, which we calculate to be 70--80
meV/Mn. Moreover, the various magnetic MnSi films we studied have
a spin polarization of carriers at the Fermi level in the range of
30--50\%.\cite{Wu:04} These results imply that the ultrathin MnSi
film on Si(001) is a candidate for magnetoelectronic materials.

For the (Si-Fe)/Si(001) film, our calculations find the AFM state
to be unstable and to converge to the FM ground state (with a very
small spin moment, as seen in Tables~\ref{TableIII} and
\ref{TableIV}). However, the FM state and the NM state are
energetically degenerate, as seen in Table~\ref{TableIV}.
Similarly, the FM state of the 2(Si-Fe)/Si(001) and
3(Si-Fe)/Si(001) films has a small spin moment and almost the same
energy as the NM state, the energy difference being less than 5
meV/Fe. Therefore we conclude that the FeSi/Si(001) films are NM,
like the CsCl-like FeSi bulk, as discussed in Sec. III.A. The
NiSi/Si(001) film is also NM, as evidenced by our computational
results that both FM and AFM states converge to the NM ground
state.

In strong contrast to the NM FeSi and NiSi films on Si(001), the
CoSi films on Si(001) have a FM ground state. This is evident from
the magnetic moments reported in Table~\ref{TableIII} and from the
energetics reported in Table~\ref{TableIV}. In our calculations, a
hypothetical AFM state of (Si-Co)/Si(001) converges to a NM state
which is, however, higher in total energy than the FM ground state
by 15 meV/Co. The 3(Si-Co)/Si(001) film is also FM with a sizable
spin moment in the middle layer (well comparable with the bulk
value of 0.63 $\mu_B$/Co), unlike the {\em ferrimagnetic}
3(Si-Mn)/Si(001) film. For 3(Si-Co)/Si(001), the layered AFM state
is higher in total energy than the FM ground state by 10 meV/Co.
Moreover, our calculations find an increasing energy difference
between the FM ground state and the NM state: 15, 17 and 28 meV/Co
in the (Si-Co), 2(Si-Co) and 3(Si-Co)/Si(001) films, respectively.
We show in Fig. 4 the layer-resolved DOS of the FM 3(Si-Co)
overlayers. The Fermi level is found to be close to a minimum of
the Co $3d$ DOS. Obviously the high DOS at the Fermi level seen in
Fig.~2 for hypothetical NM CoSi has transformed into a minimum of
the FM DOS due to exchange splitting. For this reason, the FM
state is stable. Analyzing the DOS projected onto each Si overlayer,
we find a considerable spin polarization of carriers at the Fermi
level in the interior and near-interface Si overlayers, although
those Si atoms themselves possess only a tiny induced spin moment.

\begin{figure}
\centering \includegraphics[width=8cm]{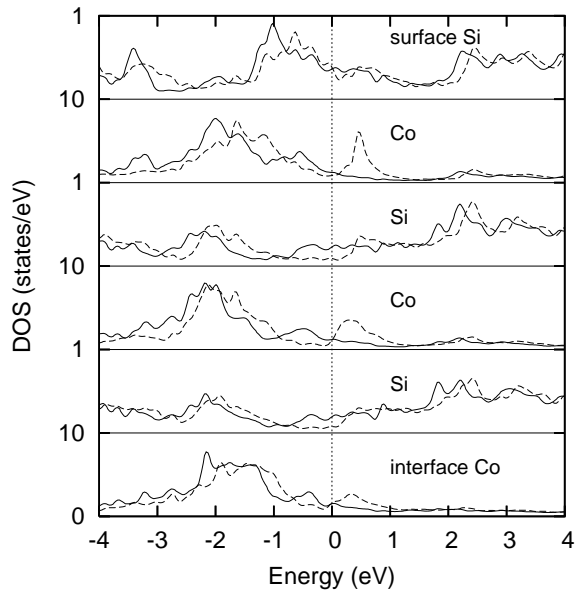} \caption{The
layer-resolved DOS of the FM 3(Si-Co)/Si(001) film.
The layers are shown from surface (top) to interface (bottom) for
the atomic structure depicted in Fig. 1 (f).
Full lines show the majority spin, dashed lines the minority
spin component.}
\end{figure}

These results suggest that the CsCl-like CoSi/Si(001) films are
interesting materials systems, having a high thermodynamic
stability among the $M$Si/Si(001) films (see Table~\ref{TableI})
and a FM metallic ground state. Since the epitaxial growth of the
CsCl-like CoSi film on Si(111) has already been
achieved,\cite{vonKanel:95} attempting to grow a CoSi/Si(001) film
may be worth the experimental effort. Moreover, the predicted
ferromagnetism of the CsCl-like CoSi calls for experimental
investigations.\cite{Note2}

\subsection{\bf $M_2$MnSi thin films on Si(001)}
\label{Heusler}
In this Section, we study films of the Heusler alloys $M_2$MnSi
($M$=Fe, Co, Ni), which one can think of as being formed by
partial Mn substitution for Si in the CsCl-like $M$Si films (cf.
Fig. 1) described so far.
In particular, the Heusler alloy Co$_2$MnSi is of
interest here, since its bulk FM half-metallicity predicted by band
calculations attracts much attention both from the
experimental\cite{Raphael:01,Ritchie:03,Singh:04} and
theoretical\cite{Kubler:83,Fujii:94,Ishida:95,
Galanakis-Dederichs:02,Picozzi:02} side. Bulk Fe$_2$MnSi, in an
ideal FM state, is also predicted by band calculations to be
half-metallic.\cite{Fujii:95} However, calculations allowing for
non-collinear alignment of the magnetic moments have found that,
in the ground state, the Mn magnetic moments are canted with
respect to the direction of the Fe magnetic moments,\cite{Mohn:98}
which leads to partial compensation of the magnetic moments along
the [111] axis. The hypothetical compound Ni$_2$MnSi, which has
not been synthesized so far to our knowledge, is shown by our
calculations {\em not} to be half metallic. For the
Co$_2$MnSi(001) surface, it has been shown recently by means of
DFT calculations\cite{Hashemifar:05} that the termination by a
Mn-Si crystal plane is thermodynamically stable, but a purely Mn-
or purely Si-terminated surface can be stable as well under very
Mn-rich or under very Si-rich conditions, respectively.

The goal of this work is to investigate how finite-size effects
and epitaxial strain in very thin films affect the magnetic
properties. The latter effect, lowering the crystallographic
symmetry, could possibly change the half-metallicity of Co$_2$MnSi
and Fe$_2$MnSi films. In particular, we investigate how possible
surface and interface electronic states affect the electronic and
magnetic properties of the films. To this end, we perform
systematic studies as a function of film thickness. Moreover, we
consider various possibilities for the surface termination of the
films, either Si surface termination [cf. Figs. 5(a) and 5(b)] or
MnSi termination [cf. Figs. 5(c), 5(d), and 5(e)]. Note that the
$M$=Fe, Co, or Ni termination is energetically unfavorable for
reasons discussed in the previous Section, and thus disregarded in
this work. In addition to the two types of surface termination,
two types of interfaces are studied, namely the $M$/Si interface
(cf. Fig. 5) and the MnSi/Si interface. The latter is
characterized by extra Mn atoms occupying the interstitial sites
of the interfacial Si layer (not shown). Firstly, we study the
$M_2$MnSi/Si(001) films with Si termination and $M$/Si interface.
Secondly, we deal with films with MnSi termination and $M$/Si
interface. Thirdly, we discuss also the MnSi/Si interface, but
restrict ourselves to Co$_2$MnSi/Si(001) films, since they are
thermodynamically stable and have a robust FM metallic ground
state, as seen below, and hence are most relevant.
\begin{figure}[th]
\centering \includegraphics[width=8cm]{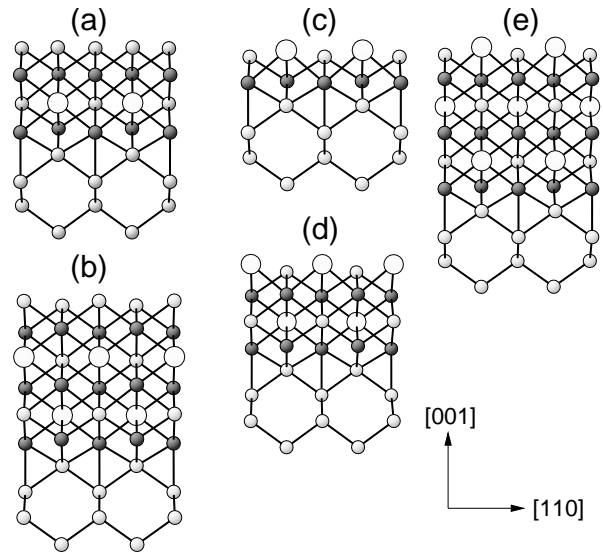}
\caption{Side view of the Si-terminated two-layered (a) and
three-layered (b) Heusler alloy $M_2$MnSi ($M$=Fe, Co, Ni) films
on Si(001) with $M$/Si interface, and of the MnSi-terminated
one-layered (c), two-layered (d), and three-layered (e) $M_2$MnSi
films. Black balls represent $M$, gray balls Si, and large white
balls Mn atoms. The bonds shorter than 2.65 ~{\AA} are shown.}
\end{figure}

\subsubsection{\bf $M_2$MnSi/Si(001): Si termination and $M$/Si interface}
In this Section, we use the terms two-layered [cf. Fig. 5(a)] and
three-layered [cf. Fig. 5(b)] Heusler alloy films, according to
the film thickness measured in repetition periods of the atomic
superstructure of the alloy. Firstly, we discuss the results for
the two- and three-layered films, focussing on magnetic ordering.
Independent on composition, we find for all the two-layered
$M_2$MnSi films a metallic ground state with FM coupling both in
the Mn sublattice and between the Mn- and $M$-sublattices
($M$=Fe,Co,Ni). For Fe$_2$MnSi, AFM ordering among the magnetic
moments of Fe and Mn is metastable, but higher than the FM state
in total energy by 20 meV per (1$\times$1) cell. For the
Co$_2$MnSi and Ni$_2$MnSi films, however, AFM ordering of the
magnetic moments of the Co and Mn (or of Ni and Mn, respectively)
is found to be unstable, and the calculations converge to the FM
ground state. Moreover, our results show that the effective Mn-Mn
FM coupling is strong, since the calculated energy cost to flip a
Mn-Mn spin pair from parallel to anti-parallel orientation is as
high as 73 meV/Mn in Fe$_2$MnSi, 216 meV/Mn in Co$_2$MnSi, and 80
meV/Mn in Ni$_2$MnSi. Note that in the two-layered $M_2$MnSi
films, the Mn atoms have the same environment as in the bulk.
Therefore it is not surprising that the calculated Mn-Mn coupling
strengths approximately scale with the measured FM Curie
temperatures of 219 K for Fe$_2$MnSi, 985 K for Co$_2$MnSi, and
320, 344, and 380~K for Ni$_2$Mn$Z$ ($Z$=Sn,Ge,Ga,
respectively).\cite{Kubler:83,Fujii:94}

Secondly, we analyze the spin magnetic moments in the films (see
Table V). On the one hand, the Mn spin moment, being generally
larger than 2 $\mu_B$, increases in the $M_2$MnSi films as $M$
varies from Fe through Co to Ni, following the same trend as in
the bulk materials. This finding can be at least partly ascribed
to decreasing $d$-$d$ hybridization among Mn and the neighboring
transition metal atoms when going from Fe to Ni, in accordance
with the increasing energy separation between the Mn $3d$ and $M$
$3d$ orbitals (see Fig.~2). On the other hand, one can argue that
the Mn spin moment in the $M_2$MnSi/Si(001) films is still smaller
than that in the $M_2$MnSi bulk. Again, this can be explained by
stronger in-plane $d$-$d$ hybridization in the film compared to
the bulk, which gives rise to more delocalized planar electronic
states and a reduced magnetic moment. The reason for this
anisotropy is that the lattice constant of bulk Si is about 4\%
smaller than that of cubic $M_2$MnSi. Hence the $M_2$MnSi films
have reduced planar lattice constant under the epitaxial
constraint. The transition metal atom $M$(=Fe,Co,Ni) has a spin
moment less than 1 $\mu_B$. In addition, the Si atom in the MnSi
layer has a small induced spin moment which is opposite to the
spin moment of the neighboring metal atom, and generally smaller
than 0.05 $\mu_B$/Si. The substrate Si atoms have an even smaller
spin moment of less than 0.02 $\mu_B$/Si oscillating in its
orientation between one substrate layer and the next one.

\begin{table*}[tb]
\caption{The layer-resolved (counted from the substrate to the surface)
atomic spin moments (in unit of $\mu_B$) of the
Si-terminated two-layered (2L) and 3L $M_2$MnSi/Si(001) films and
of the MnSi-terminated 1L, 2L and 3L $M_2$MnSi/Si(001) films (cf.
Fig. 5). All films have a $M$/Si interface.
Shown in the last three rows are the calculated atomic
spin moments of Fe$_2$MnSi and Co$_2$MnSi at the experimental
lattice constant and of Ni$_2$MnSi (not yet synthesized) at the
GGA optimized lattice constant.}
\begin{tabular} {lccccccccccc} \\
\hline\hline
Si-term.&$M$&Si4&Si3&Si2&Si1&$M$&MnSi&$M$&MnSi&$M$&Si
\\ \hline
&Fe&0.003&--0.001&0.015&--0.007&&&0.61&2.24/--0.02&0.36&0.14\\
2L&Co&0.005&0.005&0.013&--0.005&&&0.55&2.77/--0.04&0.70&0.01\\
&Ni&--0.002&--0.006&--0.002&--0.009&&&0.14&3.06/--0.04&0.13&--0.02\\ \hline
&Fe&0.001&--0&0.011&--0&0.20&2.20/--0.01&0.21&2.31/--0.01&0.35&0.08\\
3L&Co&0.004&0.005&0.007&--0.006&0.53&2.74/--0.04&0.95&2.72/--0.04&0.71&--0.01\\
&Ni&0&--0.003&0.004&--0.007&0.16&3.03/--0.03&0.28&3.14/--0.04&0.12&--0.02\\\hline\hline
MnSi-term.&$M$&Si4&Si3&Si2&Si1&$M$&MnSi&$M$&MnSi&$M$&MnSi
\\ \hline
&Fe&0.001&0&0.010&--0.005&&&&&0.84&3.42/--0.10\\
1L&Co&0&0.002&0.005&--0&&&&&0.42&3.56/--0.10\\
&Ni&--0.001&--0.002&0.001&--0.007&&&&&0.02&3.58/--0.10\\ \hline
&Fe&0.002&--0.002&0.018&--0.010&&&0.64&2.09/--0.02&--0.06&3.45/--0.10\\
2L&Co&0.005&0.004&0.013&--0.011&&&0.54&2.65/--0.05&0.82&3.52/--0.12\\
&Ni&--0.001&--0.004&--0.001&--0.005&&&0.18&3.05/--0.03&0.23&3.63/--0.10\\ \hline
&Fe&0.001&--0.002&0.011&--0.007&0.47&2.21/--0.02&0.01&2.17/--0&0.18&3.50/--0.11\\
3L&Co&0.004&0.003&0.008&--0.013&0.52&2.70/--0.04&0.99&2.73/--0.04&0.86&3.53/--0.11\\
&Ni&0&--0.003&0.002&--0.006&0.15&3.06/--0.03&0.29&3.12/--0.04&0.17&3.61/--0.11\\ \hline\hline
bulk $M_2$MnSi&$M$&Mn&Si \\ \hline
Fe$_2$MnSi&0.083&2.769&--0 \\
Co$_2$MnSi&0.987&3.013&--0.039 \\
Ni$_2$MnSi&0.290&3.330&--0.028 \\ \hline\hline
\end{tabular}
\end{table*}

\begin{figure}
\centering \includegraphics[width=7cm]{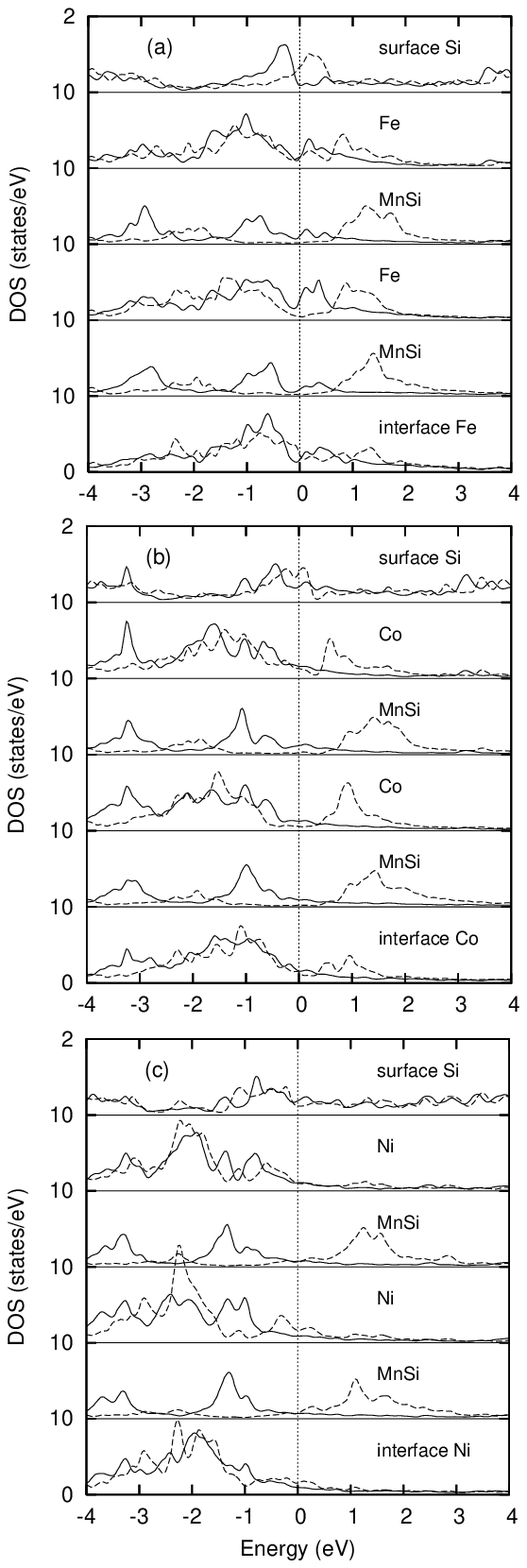}
\caption{The
layer-resolved DOS of the Si-terminated three-layered Fe$_2$MnSi
(a), Co$_2$MnSi (b) and Ni$_2$MnSi (c) films on Si(001) with
$M$/Si ($M$=Fe, Co, or Ni) interface. In each panel, the
overlayers are shown from surface (top) to interface (bottom) for
the atomic structure depicted in Fig. 5(b). Full lines show the majority
spin, dashed lines the minority spin component.}
\end{figure}

\begin{figure}
for preprint
\centering \includegraphics[width=7cm]{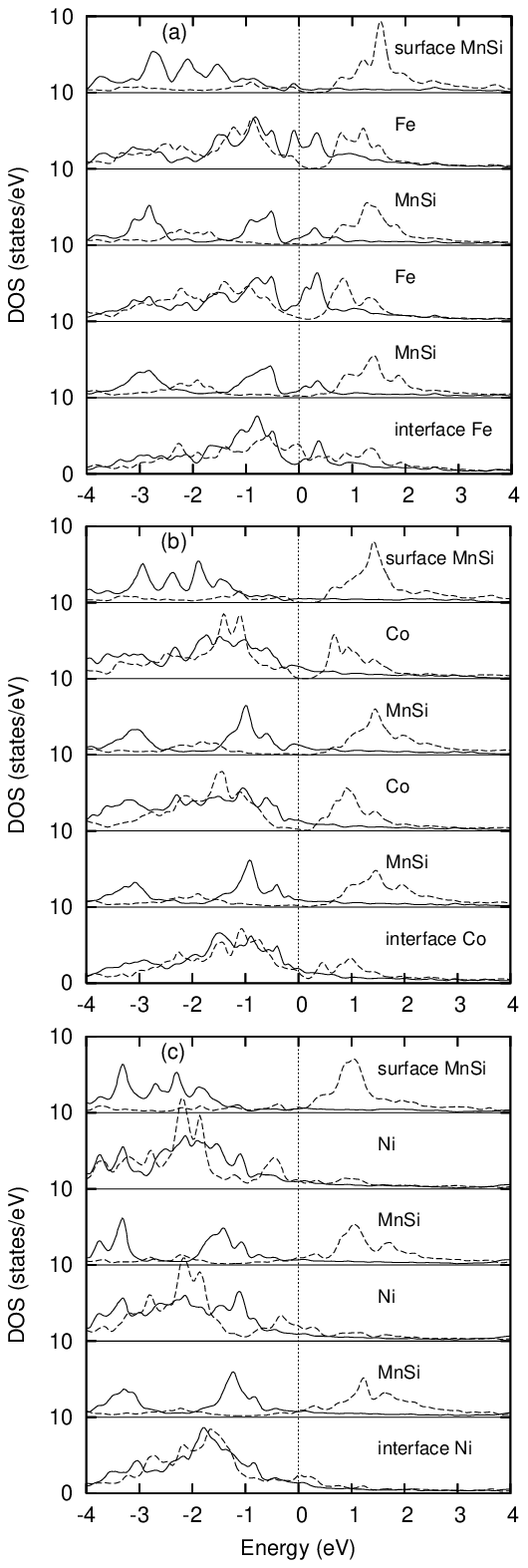}
\caption{The
layer-resolved DOS of the MnSi-terminated three-layered Fe$_2$MnSi
(a), Co$_2$MnSi (b) and Ni$_2$MnSi (c) films on Si(001) with
$M$/Si ($M$=Fe, Co, or Ni) interface. In each panel, the
overlayers are shown from surface (top) to interface (bottom) for
the atomic structure depicted in Fig. 5(e). Full lines show the majority
spin, dashed lines the minority spin component.}
\end{figure}

For the Si-terminated three-layered $M_2$MnSi films, our
calculations find, in complete analogy to the above two-layer
case, a FM metallic ground state irrespective of the nature of the
transition metal. Besides the strong FM Mn-Mn intralayer coupling
discussed above, the interlayer Mn-Mn coupling (evaluated by
switching the relative orientation of the magnetic moment in two
neighboring MnSi layers in the supercell) is 4 meV/Mn in the
Fe$_2$MnSi film, 167 meV/Mn in Co$_2$MnSi, and 30 meV/Mn in
Ni$_2$MnSi. The reduced interlayer coupling can be at least partly
ascribed to a tetragonal distortion, by noting that the Heusler
alloy film is under compressive epitaxial strain on Si(001), as
stated above, and thus has an enlarged spacing between layers. In
addition, the $M$ spin, which mediates the effective Mn-Mn
coupling, plays an important role for the magnetic ordering. Note
that in the three-layered $M_2$MnSi films, the $M$ atoms in the
layer sandwiched between two MnSi layers have an averaged spin
moment of 0.21 $\mu_B$/Fe, 0.95 $\mu_B$/Co, and 0.28 $\mu_B$/Ni,
as seen from Table V. In contrast to this, we observe that for the
layered AFM ordering of the Mn spins, the Co spin in the middle
layer is quenched to a value close to zero. The vanishing of the
Co spin moment in the layered AFM state, sitting between two
spin-antiparallel MnSi layers, is simply a consequence of
symmetry. The highest energy cost of switching from FM to AFM
alignment of the Mn spins correlates with the largest magnetic
moment at Co in the FM state in the three Heusler alloys studied
here. This indicates that the quenching of the Co spin moment is
energetically unfavourable and hence the FM state is preferred
over the AFM state.

Next, we investigate if the half-metallic properties of the
Co$_2$MnSi and Fe$_2$MnSi bulk materials also show up in the thin
films. In Fig.~6, the overlayer-resolved DOS of the Si-terminated
three-layered $M_2$MnSi ($M$=Fe,Co,Ni) films on Si(001) is shown.
Generally, the films do not show a gap in the DOS at the Fermi
level. However, the spin-polarization at the Fermi level is high
in the three middle layers, MnSi-Fe-MnSi or MnSi-Co-MnSi. We
interpret this as an incipient recovery of  the half-metallicity
of the bulk Fe$_2$MnSi and Co$_2$MnSi. However, in the Ni$_2$MnSi
film, this is not the case, consistent with our finding that bulk
Ni$_2$MnSi is {\em not} half-metallic. In all the $M_2$MnSi films
studied here, the surface Si layer has a sizable
spin-polarization ($>$30\%) at the Fermi level, following the definition
in Ref. 5, while the subsurface
$M$ and the interfacial $M$ layers have only low
spin-polarization ($<$10\%) at the
Fermi level (except for $\sim$20\% for the interfacial Ni layer).

Finally, we turn to the subject of thermodynamic stability. By
calculating the formation energy using Eq. (1), we conclude that
all Si-terminated two- and three-layered $M_2$MnSi films on
Si(001) are stable against a decomposition into the clean Si(001)
surface and bulk TMs. This is indicated by their negative {\Ef}
values, as seen in Table VI. Moreover, we checked the stability of
the $M_2$MnSi films against separated $M$Si and MnSi films by
calculating the heat of reaction, $\Delta$E, defined by
\begin{eqnarray}
\lefteqn{\textrm{$M$Si/Si(001)+MnSi/Si(001)}} & & \nonumber \\
& \rightarrow & \textrm{$M_2$MnSi/Si(001)+clean Si(001)+$\Delta$E}
\end{eqnarray}
The $M_2$MnSi films is stable (unstable) if $\Delta$E is positive
(negative). As shown by our results summarized in Table VI, the
two-layered Fe$_2$MnSi film [$\Delta$E=0.02 eV per (1$\times$1)
cell] is close to becoming unstable, and the three-layered one
[$\Delta E = -0.65$~eV per (1$\times$1) cell] is obviously
unstable. The two-layered Co$_2$MnSi film is stable while the
three-layered one tends to be unstable. However, the Ni$_2$MnSi
film is stable against a phase separation into the NiSi and MnSi
films. This is because the NiSi film is less stable due to its
oversaturated eight-fold Si coordination of Ni, while the
Ni$_2$MnSi film is stable, involving only four-fold Si
coordination of Ni.

\begin{table}[htb]
\caption{Formation energies (Eq. 1) and heat of reaction $\Delta
E$ (Eq. 2) [in unit of eV per (1$\times$1) cell] of the
Si-terminated two-layered (2L) and 3L $M_2$MnSi/Si(001) films and
of the MnSi-terminated 1L, 2L and 3L $M_2$MnSi/Si(001) films (cf.
Fig. 5). All films have a $M$/Si interface.}
\begin{tabular} {lcccccccc} \\
\hline\hline
&&&&\multicolumn{2}{c}{Si-term.}&&\multicolumn{2}{c}{MnSi-term.} \\
&&$M$&&{\Ef}&$\Delta$E&&{\Ef}&$\Delta$E \\ \hline
&&Fe&&&&&--0.20&0.86 \\
1L&&Co&&&&&--0.71&0.92 \\
&&Ni&&&&&--0.80&0.81 \\ \hline
&&Fe&&--1.08&0.02&&--1.42&0.33 \\
2L&&Co&&--1.87&0.14&&--2.30&0.53 \\
&&Ni&&--2.07&0.40&&--2.37&0.62 \\ \hline
&&Fe&&--2.51&--0.65&&--2.87&--0.66 \\
3L&&Co&&--3.48&--0.05&&--3.99&0.09 \\
&&Ni&&--3.42&0.58&&--3.69&0.48 \\ \hline\hline
\end{tabular}
\end{table}

\subsubsection{\bf $M_2$MnSi/Si(001): MnSi termination and $M$/Si
interface}

Next we deal with the $M_2$MnSi/Si(001) thin films with MnSi
termination [cf. Figs. 5(c), 5(d) and 5(e)]. The surface Mn atom
has an increased spin moment of about 3.5~$\mu_B$, and the surface
Si atom also has an increased induced spin moment of about $-0.1
\mu_B$, as seen in Table V. The spin moments of Mn and Si in the
sandwich layer between two $M$ layers are, due to the identical
environment, very similar to those in the Si-terminated $M_2$MnSi
films discussed above. The spin moment of the $M$ atom sandwiching
two MnSi layers, which plays an important role in the effective
Mn-Mn coupling, is less than 0.2 $\mu_B$/Fe, about
0.8-1.0~$\mu_B$/Co or 0.2-0.3~$\mu_B$/Ni. These values agree
closely with those of the Si-terminated three-layered $M_2$MnSi
films discussed above, and of the bulk materials. The MnSi
termination brings about a gain in the formation energy in the
range of 0.3-0.5~eV per (1$\times$1) cell for the two- and
three-layered $M_2$MnSi films (the exact value being
materials-dependent) compared with the Si-terminated $M_2$MnSi
films, which means that the former has higher thermodynamic
stability. However, we would like to draw the reader's attention
to the fact that the cohesive energy of Si is larger than that of
Mn by about 1.5~eV, as indicated by experiments and our
calculations. Combining the calculated values for the stability of
both the films and the bulk phases, we conclude that the MnSi
termination has highest thermodynamic stability mostly due to the low
cohesive energy of Mn bulk. However, Si has a higher surface
adsorption energy in the Si termination than Mn in the MnSi
termination by about 1.0~eV. In this sense, the Si-terminated
$M_2$MnSi films have stronger surface Si-$M$ bonds than the Mn-$M$
bonds present in the MnSi termination, and therefore the Si
termination is chemically more stable. Moreover, as seen in Table
VI, all the MnSi-terminated $M_2$MnSi films are stable against a
phase separation, except for the three-layered Fe$_2$MnSi film.

In Fig.~7, the overlayer-resolved DOS of the MnSi-terminated
three-layered $M_2$MnSi films are shown. The surface MnSi layer of
the Fe$_2$MnSi film brings about a notable change for the
subsurface Fe layer compared to the Si termination, as seen in
Fig. 6(a), and this Fe layer now becomes highly spin-polarized ($\sim$65\%)
at the Fermi level. The three middle layers, MnSi-Fe-MnSi, are less
affected. Again, we observe a tendency to recover the bulk
half-metallicity. In addition, the interfacial Fe layer has a
considerable spin-polarization ($\sim$45\%) at the Fermi level.
Similar changes
occur in the MnSi-terminated Co$_2$MnSi films. In particular, the
surface MnSi layer and the other overlayers, except for the
interfacial layer, become almost half-metallic. However, for the
Ni$_2$MnSi films, the surface MnSi layer brings no pronounced
changes as compared with the Si termination.

\begin{figure}
\centering \includegraphics[width=7.5cm]{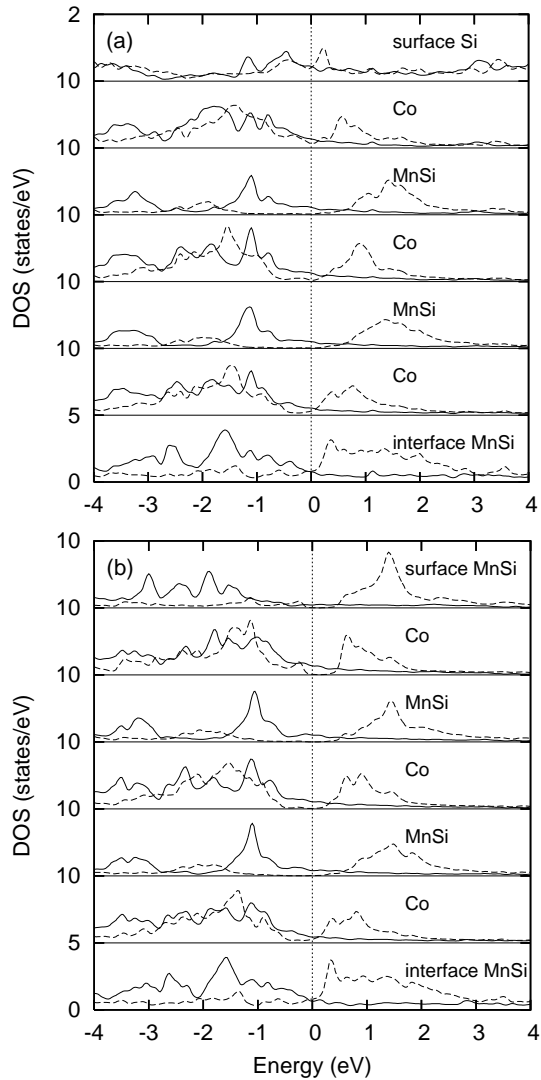} \caption{The
layer-resolved DOS of the Si-terminated (a) or MnSi-terminated (b)
three-layered Co$_2$MnSi films on Si(001) with MnSi/Si interface.
In each panel, the layers are shown from surface (top)
to interface (bottom). Full lines show the majority spin,
dashed lines the minority spin component.}
\end{figure}

\subsubsection{\bf $Co_2$MnSi/Si(001): MnSi/Si interface}
When Mn atoms occupy the interstitial sites of the interfacial Si
layer, as seen in Fig. 5, this layer now becomes a MnSi/Si
interface, replacing the former Co/Si interface. Here we
investigate Co$_2$MnSi/Si(001) films with this interface,
considering two different surface terminations, either pure Si- or
MnSi-termination. As seen in Table VII, the interfacial MnSi layer
enhances the spin moments of the overlayers, especially of the
near-interface Co layer, as compared with the Co$_2$MnSi/Si(001)
film with the Co/Si interface (Table V). Comparing films with 
the same number of Co atoms, we find that the MnSi interface 
makes the films
slightly more stable, through lowering the formation energy by 0.2 eV
per (1$\times$1)
cell or less for 1L, 2L, or 3L thickness (see Tables VI and VII
for comparison), as
a result of the low cohesive energy of bulk Mn which favors
incorporation of extra Mn atoms.
However, the Co/Si and MnSi/Si interfaces of the Co$_2$MnSi/Si(001) film
differ by less than 0.2 eV,
implying that chemical disorder in the interface layer could
occur easily through thermal fluctuations. In addition, MnSi
termination goes along with a gain in formation energy, compared
with Si termination, about 0.5 eV per (1$\times$1) cell for the
one-, two- and three-layered Co$_2$MnSi films with MnSi/Si
interface, following the same trend as in the Co$_2$MnSi/Si(001)
film with the Co/Si interface.
In Fig.~8, the overlayer-resolved DOS of both the Si- and the
MnSi-terminated three-layered Co$_2$MnSi/Si(001) films with
MnSi/Si interface are shown. Although the interfacial Mn atom has
almost the same spin moment as the middle MnSi layers where bulk
half-metallicity is almost recovered, we observe that the spin
polarization at the Fermi level in the interface layer is still
tiny ($<$10\%). Hence, in this respect, the MnSi/Si interface brings no
pronounced change for the overlayers as compared to the
Co$_2$MnSi/Si(001) film with Co/Si interface.

\begin{table*}[htb]
\caption{Formation energies [eV per (1$\times$1) cell] either of
the Si- or MnSi-terminated Co$_2$MnSi/Si(001) films with a MnSi/Si
interface (cf. Fig. 5 but note that extra Mn atoms occupy the
interstitial sites of the interfacial Si layer). The film thickness
(1L, 2L, and 3L) refers to the number of the Co-MnSi bilayers.
The third column shows the heat of reaction  $\Delta E$
[eV per (1$\times$1) cell], as defined in the text, Eq. 2. From
the fourth column onwards, the overlayer-resolved (counted from
the interface to the surface) atomic spin moments (in unit of
$\mu_B$) are shown. The substrate Si layers, each with an induced
spin moment being generally less than 0.04 $\mu_B$/Si, are
omitted.}
\begin{tabular} {lccccccccccc} \\ \hline\hline
Si-term.&{\Ef}&$\Delta$E&MnSi&Co&MnSi&Co&MnSi&Co&Si \\ \hline
1L&--0.36&0.57&&&&&2.68/--0.02&0.78&0.01\\
2L&--1.98&0.21&&&2.87/--0.01&0.98&2.79/--0.04&0.74&0.02\\
3L&--3.54&--0.36&2.77/--0.01&1.04&2.82/--0.04&1.02&2.83/--0.04
&0.81&0.03\\ \hline\hline
MnSi-term.&{\Ef}&$\Delta$E&MnSi&Co&MnSi&Co&MnSi&Co&MnSi \\ \hline
1L&--0.92&1.09&&&&&2.74/--0.02&0.88&3.54/--0.10\\
2L&--2.48&0.35&&&2.80/--0.01&1.02&2.78/--0.05&0.87&3.53/--0.11\\
3L&--4.09&--0.49&2.78/--0.01&1.03&2.82/--0.04&1.06&2.78/--0.04&0.90
&3.53/--0.11\\ \hline\hline
\end{tabular}
\end{table*}

\section{\bf Conclusion}
In summary, we have presented systematic DFT-GGA calculations for
pseudomorphic thin films of mono-silicides $M$Si ($M$=Mn, Fe, Co,
Ni) with CsCl-like atomic structure, and for thin films of Heusler
alloys $M_2$MnSi ($M$=Fe, Co, Ni) on Si(001), with particular
focus on the trends within the transition metal series.

Our calculations show that for pseudomorphic $M$Si films on Si(001),
Si surface termination is energetically preferred because it
optimizes the surface valence bond structure, i.e., four-fold
coordination of surface Si and seven- or eight-fold coordination
of subsurface $M$ atoms are achieved. The $M$-Si chemical bond
becomes stronger as $M$ varies from Mn through Fe and Co to Ni,
due to decreasing $M$ $3d$--Si $3s3p$ energy separation, and hence
increasing hybridization of the metal $3d$-states with the Si
valence band. The calculated variations in thermodynamic stability
of the $M$Si/Si(001) films can be accounted for in terms of both
the $M$ $3d$--Si $3s3p$ energy separation and the $M$ $3d$ orbital
occupation.

These trends for the bond strength also enable us to rationalize the
observed atomic ordering in Heusler alloys and to explain the
experimentally observed site preference of transition metal impurities
added to Heusler alloys. We confirm previous work\cite{Profeta:05}  
showing that CoSi films, in addition 
to ultrathin FM MnSi films\cite{Wu:04},
are another possibility to grow thin FM silicide films on Si(001),
while FeSi and NiSi films are found to be non-magnetic.
Therefore, MnSi and CoSi films on Si(001) deserve
further experimental studies.

For the $M_2$MnSi/Si(001) films, our results show that MnSi
termination is thermodynamically stable.
The slightly less stable Si termination, once formed,
is long-lived, since removing Si atoms is energetically more costly
than removing Mn atoms. Except for the atoms in the surface and
interface layers, we find that the electronic structure known from
the bulk samples is recovered quickly in the interior of the
overlayers. In particular, the half-metallicity of bulk Fe$_2$MnSi
and Co$_2$MnSi is almost recovered in the three middle layers of
the films investigated. As far as magnetic ordering in the
$M_2$MnSi films is concerned, we find that the effective
intralayer Mn-Mn FM couplings mediated by the first-neighbor $M$
atoms are strong and approximately scale with the measured Curie
temperatures of the corresponding bulk $M_2$MnSi samples. The
interlayer Mn-Mn FM coupling remains strong in the Co$_2$MnSi
films while it is (much) reduced in the Ni$_2$MnSi (Fe$_2$MnSi)
films. The Co$_2$MnSi/Si(001) thin film is thermodynamically
stable and has a robust FM metallic ground state,
and thus is most relevant for possible applications.
However, by analyzing our calculations we also
identify two effects that could possibly be detrimental for use of
these films for spin injection: The Co/Si and MnSi/Si interfaces
are found to have a similar formation energy, which makes
thermally induced interfacial disorder likely; and the
interfacial Co or MnSi layer doesn't display the gap in the
layer-resolved DOS of the minority spin channel characteristic for
a half-metal.

This work was supported by the Deutsche Forschungsgemeinschaft
through SFB 290.


\begin{thebibliography}{15}
\bibitem{Note} Using of the same $RK_{max}$=7.8 as in
our previous work\cite{Wu:04} but a little smaller $R_{\rm
Si}$=$R_{\rm Fe}$=$R_{\rm Co}$=$R_{\rm Ni}$=1.06~{\AA} (2.0~bohr)
yields the higher cut-off energy of 15.2 Ryd, compared with that
of 13.8 Ryd used previously. Consistence between our presently
calculated results of the Mn/Si(001) system and those previously
obtained, as seen in Table~\protect\ref{TableI}, indicates that
the technical settings used here ensure reliable numerical
accuracy of our results.
\end{thebibliography}
\end{document}